\documentclass[%
 reprint,
 superscriptaddress,
 amsmath,amssymb,
 aps,
 prl,
 longbibliography,
lengthcheck,%
]{revtex4-1}

\usepackage{graphicx}% Include figure files
\usepackage{dcolumn}% Align table columns on decimal point
\usepackage{bm}% bold math
\usepackage{hyperref}% add hypertext capabilities
\usepackage{color}
\usepackage{ulem}
\usepackage{xr}
\begin{document}

\preprint{APS/123-QED}

\title{
  Rectification of Spin Current in Inversion-Asymmetric Magnets\\
  with Linearly-Polarized Electromagnetic Waves 
}

\author{Hiroaki Ishizuka}
\affiliation{
Department of Applied Physics, The University of Tokyo, Bunkyo, Tokyo, 113-8656, JAPAN 
}

\author{Masahiro Sato}
\affiliation{
Department of Physics, Ibaraki University,  Mito, Ibaraki, 310-8512, JAPAN
}

\date{\today}

\begin{abstract} 
We theoretically propose a method of rectifying spin current with a linearly-polarized electromagnetic wave in inversion-asymmetric magnetic insulators. To demonstrate the proposal, we consider quantum spin chains as a simple example; these models are mapped to fermion (spinon) models via Jordan-Wigner transformation. Using a nonlinear response theory, we find that a dc spin current is generated by the linearly-polarized waves. The spin current shows rich anisotropic behavior depending on the direction of the electromagnetic wave. This is a manifestation of the rich interplay between spins and the waves; inverse Dzyaloshinskii-Moriya, Zeeman, and magnetostriction couplings lead to different behaviors of the spin current. The resultant spin current is insensitive to the relaxation time of spinons, a property of which potentially benefits a long-distance propagation of the spin current. An estimate of the required electromagnetic wave is given. 
\end{abstract}

\maketitle

{\it Introduction} --- Manipulation of magnetic states and spin current is a key subject in spintronics~\cite{Spincurrent2012}. In conductive materials, the charge current is often used for such purposes; magnetic domain walls are moved by spin-transfer effect~\cite{Berger1984}, and spin Hall effects are used to generate spin current~\cite{Dyakonov1971,Hirsch1999,Murakami2003,Sinova2004}. The concept of spintronics is also applied to magnetic insulators. They have several advantages over the metallic materials: Magnetic excitations typically have longer life time and no ohmic loss. In these magnets, the electromagnetic wave is a ``utility tool''. Recent studies demonstrate that magnetic states and excitations can be controlled by electromagnetic waves. For instance, laser control of magnetizations~\cite{Kimel2005,Stanciu2007,Kirilyuk2010,Mukai2016,Takayoshi2014-1,Takayoshi2014-2}, magnetic interactions~\cite{Mentink2015}, and magnetic textures~\cite{Mochizuki2010,Koshibae2014,Sato2016,Fujita2017-1,Fujita2017-2,Ono2017}, spin-wave propagation by focused light~\cite{Satoh2012,Hashimoto2017}, etc. have been extensively studied both experimentally and theoretically. These studies demonstrated that the electromagnetic wave has a high potentiality of controlling the magnetic states and opened a subfield utilizing lights, called opto-spintronics~\cite{Kirilyuk2010,Nemec2018}.

In contrast, the manipulation of the spin current carried by magnetic excitations is limited to ferromagnets; spin pumping with the electromagnetic wave is often used to generate the spin current~\cite{Kajiwara2010,Heinrich2011,Ohnuma2014}. On the other hand, other magnetic states (antiferromagnetic, spiral, spin liquid states, etc.) potentially have different advantages over ferromagnets. Therefore, a method for the generation and manipulation of spin current in these materials opens up an interesting possibility for spintronics. For this purpose, the usage of electromagnetic waves is desirable because of the highly precise and ultra-fast control.

%%%%%%%%%%%%%%%%%%%%%%%%%%%%%%%%%%%%%

\begin{figure}
  \includegraphics[width=\linewidth]{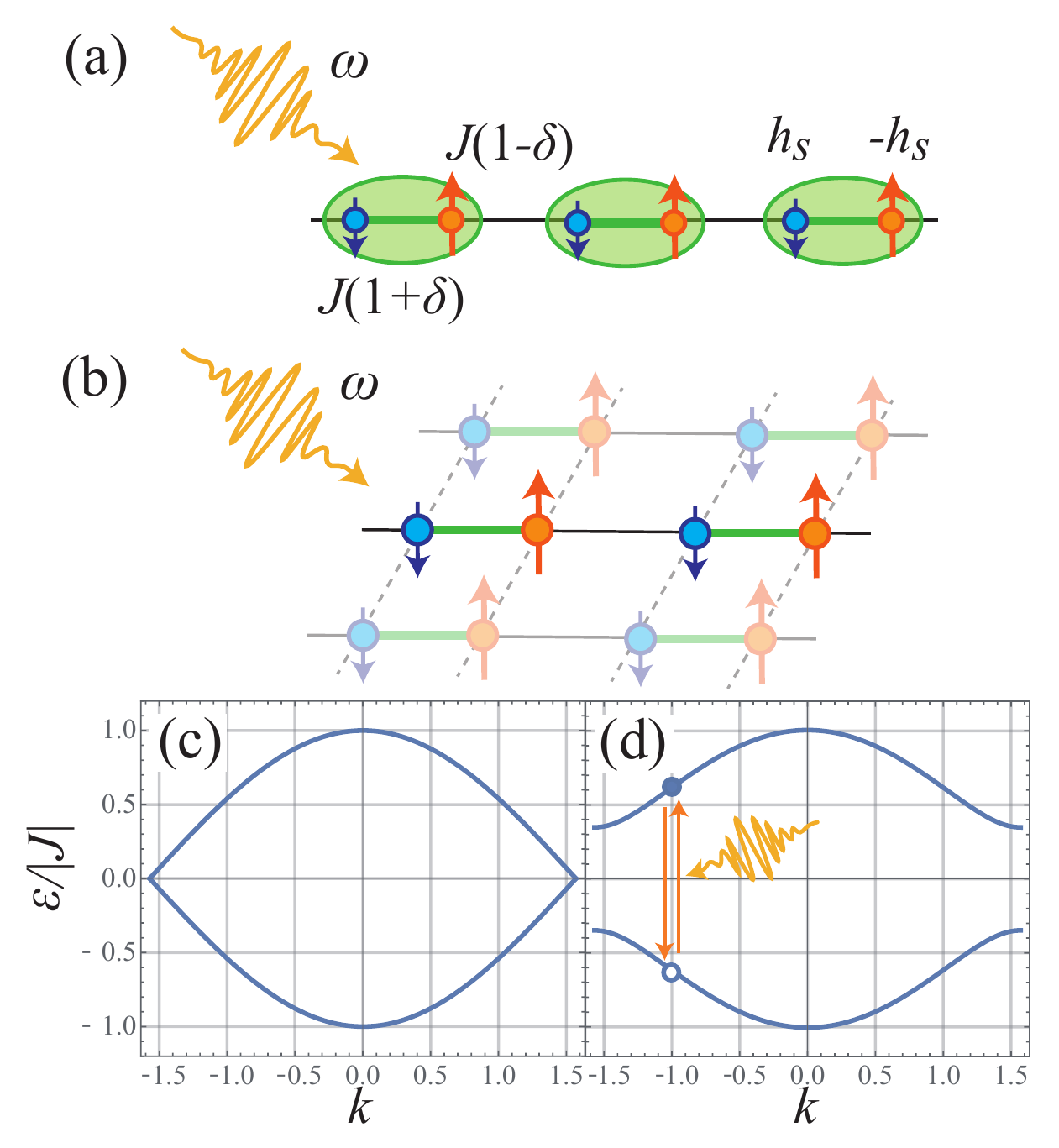}
  \caption{(Color online) Noncentrosymmetric spin chains considered in this work. Schematic picture of (a) a dimerized spin chain and (b) an antiferromagnet of weakly coupled spin chains in a staggered magnetic field. The model consists of two magnetic atoms with different $g$ factor and alternating bonds. Band structure of Jordan-Wigner fermions for (c) $\delta=h_s/J=0$, and (d) $\delta=1/3$, and $h_s/J=1/10$.}
  \label{fig:model}
\end{figure}

%%%%%%%%%%%%%%%%%%%%%%%%%%%%%%%%%%%%%%%%%%

A main issue, however, lies in moving the magnetic excitations using the electromagnetic field; the magnetic excitations do not accelerate/drift by the electromagnetic field because they are chargeless. This problem is potentially solved by utilizing the nonlinear response of magnetic insulators [Fig.~\ref{fig:model}(a,b)]. In the nonlinear optics of noncentrosymmetric electron systems~\cite{Sturman1992,Tan2016,Tokura2018}, a non-trivial dynamics of electrons during the transition process induce a ``shift'' of the particle position~\cite{Belinicher1982,Sipe2000,Morimoto2016,Ishizuka2017c}. Recent experiments investigating this mechanism find the current propagates faster than the quasi-particle velocity~\cite{Nakamura2017,Ogawa2017}. In addition, it is insensitive to the quasi-particle relaxation time; this is a beneficial property considering the heating by the electromagnetic waves reduces the relaxation time. A spin current with such interesting properties is potentially possible if the shift mechanism of magnetic excitations is generated by the electromagnetic waves.

To investigate the control of spin current by the nonlinear response, we here explore the generation of spin current by the shift mechanism in a quantum spin chain model [Fig.~\ref{fig:model}(a)]. We show that the spin current is indeed generated by simply applying a linearly-polarized electromagnetic wave if the system possesses one of the three kinds of spin-light couplings: inverse Dzyaloshinskii-Moriya (DM), Zeeman, and magnetostriction couplings. These couplings give rise to rich features in the frequency dependence and anisotropy. Interestingly, the spin current is generated by a different transition process from the electronic photogalvanic effect. The estimate of the magnitude of spin current shows our proposal gives an observable spin current with a reasonable strength of electromagnetic wave.

%%%%%%%%%%%%%%%%%%%%%%%%%%%%%%%%%%%%%

{\it Noncentrosymmetric spin chains} --- An $S=1/2$ spin chain with staggered exchange and the magnetic field is used to study the photovoltaic effect of spin current. The Hamiltonian reads
\begin{align}
  H=&\sum_i J(1+(-1)^i\delta)(S_i^xS_{i+1}^x+S_i^yS_{i+1}^y)\nonumber\\
  &\qquad-\sum_i (h+(-1)^ih_s) S_i^z.\label{eq:model}
\end{align}
Here, $S_i^{x,y,z}$ are $S=1/2$ spin operators on site $i$, $J$ is the exchange coupling whose energy scale is usually 
in gigahertz (GHz) or terahertz (THz) regime, $h$ is the uniform magnetic field along $z$ axis, 
and $h_s$ is the staggered magnetic field. This model has a wide range of applications. 
An obvious application is to the one-dimensional (1D) dimerized XY spin chains with two alternating ions [Fig.~\ref{fig:model}(a)]. 
In this case, the staggered magnetic field $h_s$ appears as a consequence of different $g$ factors for the odd- and even-site spins~\cite{Dender1997,Affleck1999,Oshikawa1999,Feyerherm2000,Morisaki2007,Umegaki2009}. 
The model can also be viewed as the effective model for a N\'eel ordered Ising-like spin chain~\cite{Shiba1980,Coldea2010,Faure2018} at zero temperature $T=0$ under a staggered magnetic field, in which the Ising interaction $J_z S_i^z S_{i+1}^z$ is treated via the mean-field approximation $S_i^z= \langle S_i\rangle+(S_i^z-\langle S_i\rangle)$. For the N\'eel ordered state, the field $(-1)^i h_s$ is the sum of the external staggered field and the mean field $J_z \langle S_i\rangle=(-1)^i J_z M_s$ ($M_s$ is the staggered magnetization). 
Furthermore, Eq.~(\ref{eq:model}) can also be applied to three-dimensional antiferromagnets of weakly coupled spin chains under a staggered field [Fig.~\ref{fig:model}(b)]. 
Treating the inter-chain coupling by a mean-field theory~\cite{Scalapino1975,Schulz1996,Sato2004,Okunishi2007} 
gives an effective one-dimensional (1D) model, Eq.~\eqref{eq:model}. 
Namely, in this system, the staggered field $h_s$ is renormalized by the inter-chain N\'eel order. 
Note that the dimerization parameter $\delta$ and the staggered field $h_s$ break site-center and bond-center inversion symmetries, respectively. Such a noncentrosymmetric nature is necessary for a photogalvanic effect.

The spin model in Eq.~\eqref{eq:model} is mapped to a fermion model using Jordan-Wigner (JW) 
transformation~\cite{Giamarchi2004,Gogolin2004,Tsvelik2007}. 
By introducing fermion operators $c_i\equiv e^{-i\pi\sum_{j=1}^{i-1}S_j^+S_j^-}S_i^-$ and 
$c_i^\dagger\equiv S_i^+ e^{i\pi\sum_{j=1}^{i-1}S_j^+S_j^-}$, Eq.~\eqref{eq:model} is fermionized as 
\begin{align}
  H=\sum_i&\frac{J(1+(-1)^i\delta)}2\left(c_{i+1}^\dagger c_i+c_{i}^\dagger c_{i+1}\right)\nonumber\\
  &+(h+(-1)^ih_s)n_i.\label{eq:modelJW}
\end{align}
Here, $S^\pm_i\equiv S_i^x\pm iS_i^y$ are the ladder operators and $n_i\equiv c_i^\dagger c_i$ is the number operator for the fermions at $i$th site.  Figures~\ref{fig:model}(c) and \ref{fig:model}(d) show the band structure of the JW fermions. The model has a band gap $\Delta_{\frac\pi2}\equiv2\sqrt{J^2\delta^2+h_s^2}$ for $|\delta|<1$ [Fig.~\ref{fig:model}(d)], while the gap is $\Delta_0\equiv2\sqrt{J^2+h_s^2}$ if $|\delta|>1$. The model is gapless only if $h_s=\delta=0$ [Fig.~\ref{fig:model}(c)]. Therefore, the ground state is robust against $h$ as long as $h<\Delta/2$, where $\Delta\equiv\min(\Delta_0,\Delta_{\frac\pi2})$. We focus on the weak $h$ region of this model in the rest of this work.

The spin current operator for $S^z$ is defined from the continuity equation. 
The current density operator reads
\begin{align}
  J_{sc} &\equiv \frac1L\sum_iJ(1+(-1)^i\delta)(S_{i+1}^xS_i^y-S_{i+1}^yS_i^x),
\end{align}
where $L$ is the number of sites; here, we set the Planck constant $\hbar=1$. 

%%%%%%%%%%%%%%%%%%%%%%%%%%%%%%%%%%%%%%%%%%%%%%%%%%%%%
{\it Inverse DM coupling} --- 
External electromagnetic waves couple to spins in several different forms. First, we consider the coupling of the electric field to the electric dipole induced by the inverse DM mechanism~\cite{Katsura2005,Takahashi2012,Huvonen2009,Furukawa2010,Tokura2014}:
\begin{align}
  H_\text{iDM} =& E_y(t)\sum_i (p+(-1)^ip_s)\left(\bm S_i\times\bm S_{i+1}\right)^z.
\end{align}
Here, the chain is along the $x$ axis, $p\mp p_s$ is the coefficient for the ferroelectric polarization of odd and even bonds, and 
$E_y(t)=E_y\cos(\omega t)$ is the oscillating electric field along the $y$ axis with frequency 
$\omega$ (typically, GHz or THz). Note that at a special point $p_s/p=\delta$, the term $H_\text{iDM}$ is analogous to the linear-order coupling of the electrons to the vector potential. We will comment on this case later. 

The spin current conductivity is calculated using a quadratic response formula similar to 
that for photovoltaic effects~\cite{Kraut1979}. The formula reads
\begin{align}
&\sigma^{(2)}(\omega)=\sum_{\alpha,\beta,\gamma}\int\frac{dk}{2\pi} \frac{[f_{\alpha}(k)-f_{\beta}(k)]B_{\alpha\beta}(k)}{\omega-\varepsilon_{\beta}(k)+\varepsilon_{\alpha}(k)-i/(2\tau)}\nonumber\\
&\times\left[\frac{B_{\beta\gamma}(k) J_{\gamma\alpha}(k)}{\varepsilon_{\alpha}(k)-\varepsilon_{\gamma}(k)-i/(2\tau)}-\frac{J_{\beta\gamma}(k)B_{\gamma\alpha}(k)}{\varepsilon_{\gamma}(k)-\varepsilon_{\beta}(k)-i/(2\tau)}\right],\label{eq:KvB}
\end{align}
where $\varepsilon_{\alpha}(k)$ is the eigenenergy of an $\alpha$th-band state with momentum $k$ ($\left|\alpha k\right>$), $f_\alpha(k)\equiv (1+e^{\varepsilon_{\alpha(k)/(k_BT)}})^{-1}$ is the fermion distribution 
for $\left|\alpha k\right>$, $\tau$ is the relaxation time of JW fermions, $J_{\alpha\beta}(k)\equiv\left<\alpha k\right|J_{sc}\left|\beta k\right>$, and $B_{\alpha\beta}(k)\equiv\left<\alpha k\right|H_\text{iDM}\left|\beta k\right>$. 
Hereafter, we will mainly consider the $T=0$ case of the model in Eq.~\eqref{eq:modelJW}. The conduction and valence bands [Fig.~\ref{fig:model}(c) and (d)] respectively correspond to $\alpha=+$ and $-$. 
We focus on the real part of $\sigma^{(2)}(\omega)$ because only the real part contributes to the spin current. With these simplifications, Eq.~\eqref{eq:KvB} becomes
\begin{align}
\Re&[\sigma^{(2)}(\omega)]=\nonumber\\
&\frac1{\pi}\Re\left\{\sum_{k}\frac{B_{+-}(k)J_{-+}(k)[B_{--}(k)-B_{++}(k)]}{\omega^2-[\varepsilon_{+,k}-\varepsilon_{-,k}-i/(2\tau)]^2}\right\},\label{eq:KvB2}
\end{align}
provided that $\varepsilon_{k\pm}$ and $|B_{+-}|^2$ are even with respect to $k$.

\begin{figure}[tb]
  \includegraphics[width=\linewidth]{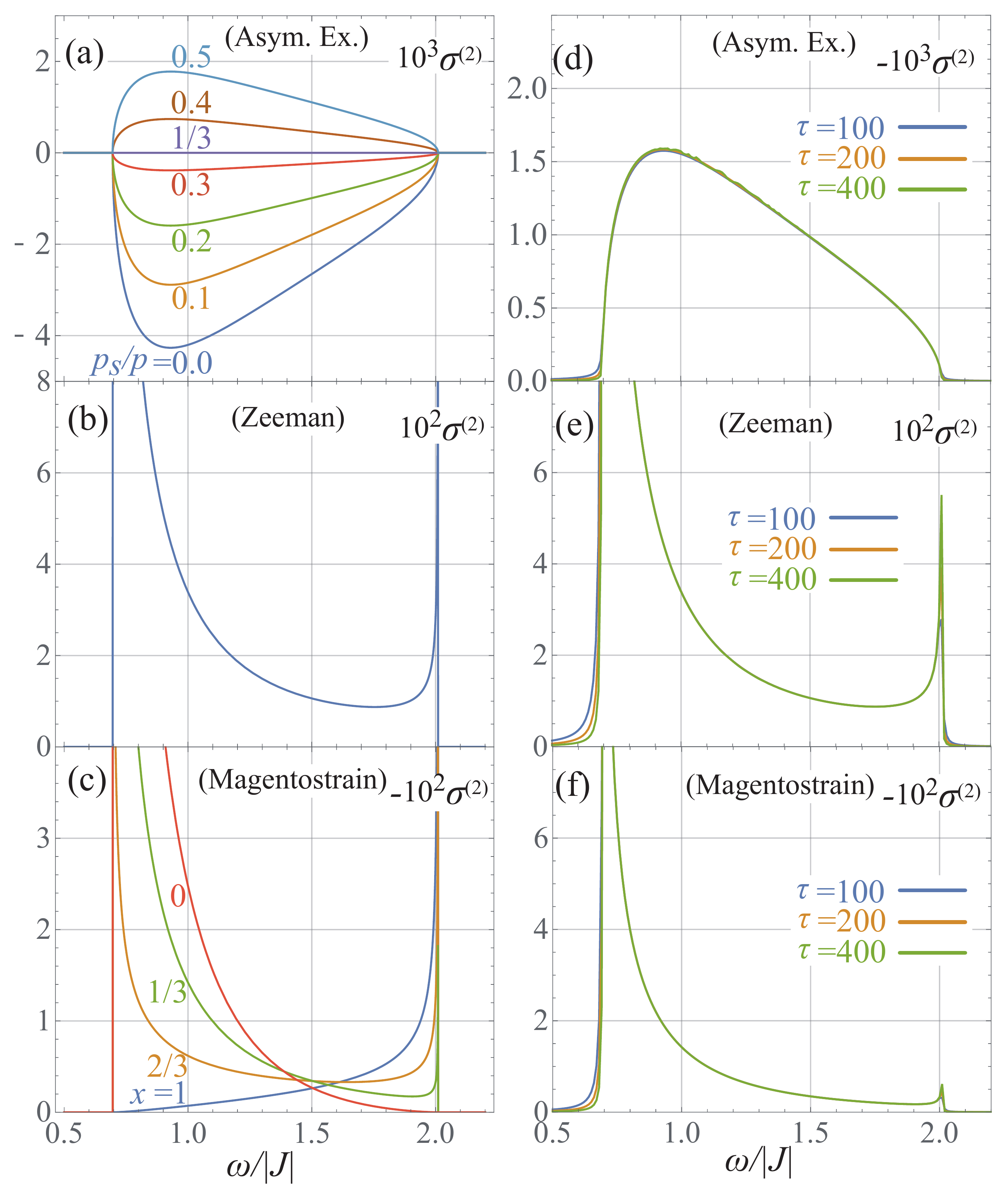}
  \caption{(Color online) Frequency dependence of the nonlinear conductivity $\sigma^{(2)}(\omega)$ for $J=1$, $\delta=1/3$, and $h_s=1/10$. Figures (a)-(c) are the results for $\tau\to\infty$ with (a) inverse Dzyaloshinskii-Moriya coupling with $p=1$, (b) Zeeman coupling, and (c) magnetostriction effect with $A=x$ and $A_s=1-x$. Different lines in (a) and (c) are the results for different ratio of parameters $p_s/p$ and $A/B$, respectively. Figures (d)-(f) shows the $\tau$ dependence of (d) inverse DM coupling with $p=1$ and $p_s=0.2$, (e) Zeeman coupling, and (f) magnetostriction effect with $A=1/3$ and $A_s=2/3$.}
  \label{fig:analytic}
\end{figure}

Using Eq.~\eqref{eq:KvB2}, the nonlinear conductivity in the $\tau \to \infty$ limit becomes
\begin{align}
\Re[\sigma^{(2)}(\omega)]={\rm sgn}&(1-\delta^2)\frac{h_s(p_s-p\delta)(p-p_s\delta)}{2\pi\omega^2J^2(1-\delta^2)^2}\nonumber\\
&\times\sqrt{(\omega^2-\Delta_{\frac\pi2}^2)(\Delta_{0}^2-\omega^2)},\label{eq:sigmaAS}
\end{align}
when $\Delta\leq\omega\leq W\equiv\max(\Delta_{0},\Delta_{\frac\pi2})$. On the other hand, no spin current appears for a frequency $\omega<\Delta$ or $W<\omega$, which implies that an inter-band optical transition is necessary for the spin current. Figure~\ref{fig:analytic}(a) shows the result for $J=1$, $\delta=1/3$, and $h_s=1/10$. The conductivity becomes zero when $h_s=0$ or $p_s=\delta=0$ and is proportional to $h_s(p_s-p\delta)$. These features reflect the symmetry property of the conductance. 
The model becomes inversion symmetric when $h_s=0$ or $p_s=\delta=0$, and therefore, the conductivity vanishes.
For the noncentrosymmetric chain, the inversion operation imposes following relations: $\sigma^{(2)}(\omega;\delta,h_s,p_s)=-\sigma^{(2)}(\omega;-\delta,h_s,-p_s)$ and $\sigma^{(2)}(\omega;\delta,h_s,p_s)=-\sigma^{(2)}(\omega;\delta,-h_s,p_s)$~\cite{suppl}. 
Hence, the lowest order terms in the symmetry-breaking parameters are proportional to $h_s\delta$ or $h_s p_s$. 
Another important feature is that the spin current vanishes when $\delta=p_s/p$. This is a well-known result in the photocurrent; the photocurrent induced by the linear-coupling terms vanishes in two-band models~\cite{Kraut1979}. In contrast, in general, a finite spin current appears in our case because $B_{\alpha\beta}(k)$ is generally different from the current operator.

%%%%%%%%%%%%%%%%%%%%%%%%%%%%%%%%%%%%%
%%%%%%%%%%%%%%%%%%%%%%%%%%%%%%%%%%%%%
%%%%%%%%%%%%%%%%%%%%%%%%%%%%%%%%%%%%%

We find that the nonlinear conductance in Eq.~\eqref{eq:sigmaAS} shows a characteristic structure when the frequency is close to $\Delta$, i.e., close to the lowest frequency with non-zero $\Re[\sigma^{(2)}(\omega)]$. 
The asymptotic form of $\Re[\sigma^{(2)}(\omega)]$ reads $\sigma^{(2)}(\omega)\propto\sqrt{\delta\omega}$, where $\delta\omega=\omega-\Delta$~\cite{suppl}. This frequency dependence is related to the momentum dependence of $g(k)\equiv B_{+-}(k)J_{-+}(k)[B_{--}(k)-B_{++}(k)]$ at the band edge. The real part of $g(k)$ is always zero in our model. Therefore, Eq.~\eqref{eq:KvB2} becomes
\begin{align}
\sigma^{(2)}(\omega)=&\frac1{8}\frac{\Im[g(k_0+k_\omega)]}{\varepsilon_+(k_0+k_\omega)}\rho[\varepsilon_+(k_0+k_\omega)],\label{eq:KvB2dos}
\end{align}
where $\rho(\varepsilon)$ is the density of states (DOS) and $k_\omega>0$ is a wavenumber such that 
$\omega=\varepsilon_+(k_0+k_\omega)-\varepsilon_-(k_0+k_\omega)$. 
Here, $k_0$ is the location of the band bottom; it is $k_0=\pi/2$ ($k_0=0$) when $1>\delta^2$ ($1<\delta^2$). 
By definition, $\delta\omega=\varepsilon_+(k_0+k_\omega)-\varepsilon_-(k_0+k_\omega)-\Delta$ and $k_\omega\to 0$ 
when $\delta\omega\to0$. 
The asymptotic form $g(k_0+k_\omega)\propto k_\omega^n$ makes 
$\sigma^{(2)}(\omega)\propto\delta\omega^{\frac{n-1}2}$ through the relations $\delta\omega\propto k_\omega^2$ 
and $\rho\propto1/\sqrt{\delta\omega}$. 
For the present case, $g(k_0+k_\omega)\propto k_\omega^2$ leads to $\sigma^{(2)}(\omega)\propto\sqrt{\delta\omega}$. 
In other words, the asymptotic form of $\sigma^{(2)}(\omega)$ reflects $g(k)$, i.e., $B_{\alpha\beta}(k)$. 
As shown below, different asymptotic form of $g(k)$ and $\sigma^{(2)}(\omega)$ appears for different kinds of spin-light couplings.

%%%%%%%%%%%%%%%%%%%%%%%%%%%%%%%%%%%%%%%%%%%%%%%%%%%%%%%%
{\it Zeeman coupling} --- The Zeeman coupling also contributes to the spin current. 
We here consider an oscillating magnetic field $B(t)=B\cos(\omega t)$ parallel to the magnetic moments. 
The Hamiltonian reads:
\begin{align}
  H_\text{Z}=&-B(t)\sum_i(\eta-(-1)^i\eta_s)S_i^z.
\label{eq:Zeeman}
\end{align}
This is in contrast to the case of usual spin pumping~\cite{Kajiwara2010,Heinrich2011,Ohnuma2014}, 
in which an oscillating magnetic field perpendicular to the magnetic moment is considered.
%We stress that the applied oscillating magnetic field is perpendicular to the magnetic 
%moment in the case of usual spin pumping~\cite{Kajiwara2010,Heinrich2011,Ohnuma2014}, 
%while the field $B(t)$ in Eq.~(\ref{eq:Zeeman}) is parallel to the static fields $h$ and 
%$h_s$.
The spin current is calculated using Eq.~\eqref{eq:KvB2} 
by the replacement $B_{\alpha\beta}(k)\to\left<\alpha k\right|H_\text{Z}\left|\beta k\right>$. The result reads
\begin{align}
\sigma^{(2)}(\omega)=\frac{8{\rm sgn}(1-\delta^2)\delta\, J^4h_s \eta_s^2}{\pi\omega^2\sqrt{\left(\omega^2-\Delta_{\frac\pi2}^2\right)\left(\Delta_{0}^2-\omega^2\right)}},\label{eq:sigma_Zeeman}
\end{align}
at $T=0$ and $\tau\to\infty$. The photocurrent depends on the staggered magnetic field $\eta_s$ and not to $\eta$. 
This follows from the form of the two-band equation in Eq.~\eqref{eq:KvB2}. Naively, three terms appear for $H_\text{Z}$, which are proportional to $\eta_s^2$, $\eta\,\eta_s$, and $\eta^2$. However, the $\eta\eta_s$ term has $B_{\alpha\beta}(k)=\eta\hat 1_{\alpha\beta}$ for one of the two $B_{\alpha\beta}(k)$'s in Eq.~\eqref{eq:KvB2} [$B_{+-}(k)$ or $B_{--}(k)-B_{++}(k)$]. As $B_{+-}(k)=B_{--}(k)-B_{++}(k)=0$ for $B_{\alpha\beta}(k)=\eta\hat 1_{\alpha\beta}$, the $\eta\,\eta_s$ term vanishes. Similarly, the $\eta^2$ term also vanishes. Hence, only the staggered magnetic field contributes to the spin current.

A notable difference from the %asymmetric exchange 
inverse DM case appears at the lower edge of the spectrum at $\omega=\Delta$. Figure~\ref{fig:analytic}(b) shows the result of $\sigma^{(2)}$ for $\eta_s=1$. The conductivity shows a divergence; the asymptotic form is $\sigma^{(2)}(\omega)\propto 1/\sqrt{\delta\omega}$~\cite{suppl}.
%\begin{align}
%\sigma^{(2)}(\omega)\approx \frac{\delta\,h_s\eta_s^2}{2\pi|1-\delta^2|^{\frac12}(J^2\delta^2+h_s^2)^{\frac54}}\frac1{\sqrt{\delta\omega}}.
%\end{align}
The divergence is a consequence of the asymptotic form of $g(k)$, which behaves differently from the asymmetric exchange case; $g(k)$ for the Zeeman coupling become a constant when $\omega\searrow\Delta$. The substitution of $g(k)$ into 
Eq.~\eqref{eq:KvB2dos} gives the asymptotic form $\sigma^{(2)}\propto\rho(k_\omega)\propto \delta\omega^{-1/2}$. Hence, the divergence reflects the structure of the DOS.

%%%%%%%%%%%%%%%%%%%%%%%%%%%%%%%%%%%%%%%%%%%%%%%%%%
{\it Magnetostriction effect} --- 
Magnetostriction effect also leads to a coupling between local exchange interaction and an external electromagnetic 
field~\cite{Tokura2014,Pimenov2006,Miyahara2008,Mochizuki2010-2,Katsura2009,Furukawa2010}; the Hamiltonian reads
\begin{align}
H_\text{ms}=E_x(t)\sum_i \{A+(-1)^iA_s\}(S_i^x S_{i+1}^x+S_i^y S_{i+1}^y).
\end{align}
Here, $A$ and $A_s$ are the uniform and staggered magnetostriction terms, respectively, 
and $E_x(t)=E_x\cos(\omega t)$ is the oscillating electric field along the $x$ axis. 
$A$ ($A_s$) is the magnetostriction effect for $J$ ($J\delta$).

The solution for $H_\text{ms}$ at $T=0$ and $\tau\to\infty$ reads
\begin{align}
\sigma^{(2)}(\omega)=&-\frac{{\rm sgn}(1-\delta^2)h_s}{4\pi\omega^2J^2(1-\delta^2)^2\sqrt{(\omega^2-\Delta_{\frac\pi2}^2)(\Delta_{0}^2-\omega^2)}}\nonumber\\
&\times\left\{A(\Delta_{\frac\pi2}^2-\omega^2)+A_s\delta(\omega^2-\Delta_{0}^2)\right\}\nonumber\\
&\times\left\{A_s(\Delta_{0}^2-\omega^2)+A\delta(\omega^2-\Delta_{\frac\pi2}^2)\right\}.\label{eq:sigma_ms}
\end{align}
Figure~\ref{fig:analytic}(c) shows the $\omega$ dependence of $\sigma^{(2)}(\omega)$ 
for $J=1$, $\delta=1/3$, and $h_s=1/10$. 
Unlike the other two cases, the asymptotic structure at $\omega\sim\Delta$ changes depending on $A$ and $A_s$. 
When $\delta^2<1$, a divergent structure similar to the Zeeman coupling, $\sigma^{(2)}(\omega)\approx\frac1{\sqrt{\delta\omega}}$, appears for $A_s\ne0$. On the other hand, 
the conductivity smoothly goes to zero at $\omega=\Delta$ for $A_s=0$; in this case, $\sigma^{(2)}(\omega)\approx\delta\omega^{\frac32}$ at the lower edge. 
Therefore, the magnetostriction effect also contributes to the spin current with a characteristic behavior at the lower edge $\omega\sim\Delta$. Further details are presented in the supplementary information~\cite{suppl}.

%%%%%%%%%%%%%%%%%%%%%%%%%%%%%%%%%%%%%%%%
{\it Relaxation time dependence} --- The $\tau$ dependence of a light-induced current often reflects its microscopic mechanism. For instance, in the study of photovoltaic effect, shift current does not depend on $\tau$ while the injection current is linearly proportional to $\tau$~\cite{Sturman1992,Sipe2000}. The numerical results of $\sigma^\text{(2)}(\omega)$ for different $\tau$ are shown in Figs.~\ref{fig:analytic}(d)-(f); each figure shows the results for (d) asymmetric exchange, (e) Zeeman, and (f) magnetostriction couplings. 
All results are calculated using $L=2^{14}$ site chains with periodic boundary condition. 
The result shows that the photo spin current is insensitive against the value of $\tau$. 
Therefore, the spin current is robust against the suppression of the relaxation time. This behavior is similar to the shift current in electronic photogalvanic effects.

%%%%%%%%%%%%%%%%%%%%%%%%%%%%%%%%%%%%%%%%%%%%%%%%%
{\it Discussion} --- In this work, we explored the generation of spin current using nonlinear response.
To this end, we considered simple but realistic quantum spin chains with three different types of couplings between spins and electromagnetic field: Inverse DM, Zeeman, and magnetostriction couplings. 
The spin current generated by all three mechanisms is independent of relaxation time of the magnetic excitation. However,  our simple model shows the spin current appears from different microscopic processes compared with the relaxation-time-independent electronic photocurrent (shift current)~\cite{Kraut1979,Belinicher1982,Sipe2000}. This feature is crucial for magnets as the total number of the bands are much less than the electronic bands. Therefore, our proposal for the spin current is generally expected in simple magnetic structures.

Another interesting feature is the anisotropy of the spin current. In our model, the spin current by inverse DM and magnetostriction couplings can be switched by rotating the electric field; the field along $y$ axis gives inverse DM component while $x$ gives the magnetostriction. Similarly, Zeeman coupling contributes when the magnetic field along $z$ axis. This anisotropy in the microscopic mechanism is reflected in the frequency dependence. Experimentally, the obseravtion of the anisotropy distinguishes the microscopic mechanism of the spin current.

We also stress that the mechanism of generating spin current differs from spin pumping~\cite{Kajiwara2010,Heinrich2011,Ohnuma2014}. Unlike the spin pumping, all three mechanisms we considered preserves the spin angular momentum along $z$ axis. Therefore, in contrast to the spin pumping, no angular momentum is supplied from the electromagnetic waves. The conservation decidedly shows that the spin current studied here is by the nontrivial motion of magnetic excitations.

In the last, we estimate the order of the spin current for each of the contributions. A typical value of exchange interactions are used for the estimate: 
$J\sim10^2k_B$ J, $\delta\sim 0.1$, and $h_s \sim \mu_B$ J. The relative permittivity $\epsilon_r=10$ is assumed. 
The excitation gap for these values reads $\Delta_g=\Delta_{\frac\pi2}\sim 1$meV. 
Therefore, the frequency of the light is assumed to be $\omega\sim\Delta_{\frac\pi2}/\hbar\sim1$ terahertz. 
Using these values, we compute the strength of the oscillating electric field 
required for the spin current density $J^{(0)}_{sc}=10^{-16}$ J/cm$^2$. 
Here $J_{sc}^{(0)}$ is an expected, typical value of the spin current observed in a recent experiment of the spin Seebeck effect for a quasi-1D magnet $\rm Sr_2CuO_3$~\cite{suppl,Hirobe2017}. 
For the inverse DM coupling, the magnitude of the electric polarization $p\sim10^{-31}$ Cm and $p_s\sim10^{-32}$ Cm are used~\cite{Jia2006,Ishizuka2018}. A bulk solid of the spin chains aligned with $\sim a_0=4$\AA\, distance gives the nonlinear conductivity $\sigma_{3D}^{(2)}\equiv\sigma^{(2)}/a_0^2\sim 10^{-25}$ A$^2$s$^4$/m$^2$kg. 
Therefore, $E_y=\epsilon_r(J_{sc}^{(0)}/\sigma_{3D}^{(2)})^{1/2}\sim 10^5$ V/cm is required 
so that the spin current approaches the value of $J_{sc}^{(0)}$. 
Similarly, in the case of Zeeman coupling, the staggered moment $\eta_s\sim0.1\mu_B$ J/T gives 
$\sigma_{3D}^{(2)}\sim10^{-9}$ J/T$^2$m$^2$. 
This requires $B\sim 10^{-2}$ T (or $E\sim 10^4$ V/cm) to induce $J_{sc}^{(0)}$. 
In the last, the magnetostriction coupling with $A\sim A_s\sim10^{-28}$ Jm/V~\cite{suppl} gives $\sigma_{3D}^{(2)}\sim10^{-19}$ A$^2$s$^4$/m$^2$kg and $E\sim10^2$ V/cm. Therefore, the spin current generated by all three mechanisms should be observable in experiments.

%%%%%%%%%%%%%%%%%%%%%%%%%%%%%%%%%%%%%%%%%%%%%%%%%%%%
This work was supported by JSPS KAKENHI (Grant Numbers JP16H06717, JP18H03676, JP18H04222, JP26103006, JP17K05513 and No. JP15H02117), ImPACT Program of Council for Science, Technology and Innovation (Cabinet office, Government of Japan), CREST JST (Grant No. JPMJCR16F1), and Grant-in-Aid for Scientific Research on Innovative Area ``Nano Spin Conversion Science”(Grant No.17H05174).

%%%%%%%%%%%%%%%%%%%%%%%%%%%%%%%%%%%%%%%%%%%%%%%%%%%%


\begin{thebibliography}{99}

%\bibitem{Moore2010}     J. E. Moore and J. Orenstein, Phys. Rev. Lett. 105, 026805 (2010).
%\bibitem{Sodemann2015}  I. Sodemann and L. Fu, Phys. Rev. Lett. 115, 216806 (2015).
%\bibitem{Ishizuka2016}  H. Ishizuka, T. Hayata, M. Ueda, and N. Nagaosa, Phys. Rev. Lett. {\bf 117}, 216601 (2016).
%\bibitem{Ishizuka2017b} H. Ishizuka, T. Hayata, M. Ueda, and N. Nagaosa, Phys. Rev. B {\bf 95}, 245211 (2017).
%\bibitem{Wu2016}         L. Wu, S. Patankar, T. Morimoto, N. L. Nair, E. Thewalt, A. Little, J. G. Analytis, J. E. Moore, and J. Orenstein, Nat. Phys. {\bf 13}, 350 (2016).
%\bibitem{Ma2017}         Q. Ma, S.-Y. Xu, C.-K. Chan, C.-L. Zhang, G. Chang, Y. Lin, W. Xie, T. Palacios, H. Lin, S. Jia, P. A. Lee, P. Jarillo-Herrero, and N. Gedik, Nat. Phys. {\bf 13}, 842 (2017).
%\bibitem{Osterhoudt2017} G. B. Osterhoudt, L. K. Diebel, X. Yang, J. Stanco, X. Huang, B. Shen, N. Ni, P. Moll, Y. Ran, and K. S. Burch, preprint (arXiv:1712.04951).
%%%%%%%%%%%%  optical spintronics  --laser control of magnetizations %%%%%%%%%%%%%%%%%%%
\bibitem{Spincurrent2012} {\it Spin Current}, edited by S. Maekawa, S. O. Valenzuela, E. Saitoh, and T. Kimura (Oxford University Press, Oxford, England, 2012).
\bibitem{Berger1984}      L. Berger, J. Appl. Phys. {\bf55}, 1954 (1984).
\bibitem{Dyakonov1971}    M. I. Dyakonov and V. I. Perel, JETP Lett. {\bf13}, 467 (1971).
\bibitem{Hirsch1999}      J. E. Hirsch, Phys. Rev. Lett. {\bf83}, 1834 (1999).
\bibitem{Murakami2003}    S. Murakami, N. Nagaosa, and S.-C. Zhang, Science {\bf301}, 1348 (2003).
\bibitem{Sinova2004}      J. Sinova, D. Culser, Q. Niu, N. A. Sinitsyn, T. Jungwirth, and A. H. MacDonald, Phys. Rev. Lett. {\bf92}, 16603 (2004).
%\bibitem{Beaurepaire1996} E. Beaurepaire, J.-C. Merle, A. Daunois, and J.-Y. Bigot, Phys. Rev. Lett. {\bf 76}, 4250 (1996).
\bibitem{Kimel2005} A. V. Kimel, A. Kirilyuk, P. A. Usachev, R. V. Pisarev, A. M. Balbashov, and T. Rasing, Nature {\bf 435}, 655 (2005).
\bibitem{Stanciu2007} C. D. Stanciu, F. Hansteen, A. V. Kimel, A. Kirilyuk, A. Tsukamoto, A. Itoh, T. Rasing, Phys. Rev. Lett. {\bf 99}, 047601 (2007). 
\bibitem{Kirilyuk2010} A. Kirilyuk, A. V. Kimel, and T. Rasing, Rev. Mod. Phys. {\bf 82}, 2731 (2010).
\bibitem{Mukai2016}Y. Mukai, H. Hirori, T. Yamamoto, H. Kageyama, and K. Tanaka, New J. Phys. {\bf 18}, 013045 (2016).
\bibitem{Takayoshi2014-1} S. Takayoshi, H. Aoki, and T. Oka, Phys. Rev. B {\bf 90}, 085150 (2014).
\bibitem{Takayoshi2014-2} S. Takayoshi, M. Sato, and T. Oka, Phys. Rev. B {\bf 90}, 214413 (2014).
%%%%%%%%%%   optical spintronics --controlling exchange interactions
\bibitem{Mentink2015}   J. H. Mentink, K. Balzer, and M. Eckstein, Nature Commun. {\bf6}, 6708 (2015).
%%%%%%%%%%   optical spintronics --controlling spin textures and magnetic orders
\bibitem{Mochizuki2010} M. Mochizuki and N. Nagaosa, Phys. Rev. Lett. {\bf 105}, 147202 (2010).
\bibitem{Koshibae2014} W. Koshibae and N. Nagaosa, Nature Commun. {\bf 5}, 5148 (2014).
\bibitem{Sato2016} M. Sato, S. Takayoshi, and T. Oka, Phys. Rev. Lett. {\bf 117}, 147202 (2016).
\bibitem{Fujita2017-1} H. Fujita and M. Sato, Phys. Rev. B {\bf 95}, 054421 (2017).
\bibitem{Fujita2017-2} H. Fujita and M. Sato, Phys. Rev. B {\bf 96}, 060407(R) (2017).
\bibitem{Ono2017} A. Ono and S. Ishihara, Phys. Rev. Lett. {\bf119}, 207202 (2017).
%%%%%%%%%%  laser-driven spin-wave
\bibitem{Satoh2012} T. Satoh, Y. Terui, R. Moriya, B. A. Ivanov, K. Ando, E. Saitoh, T. Shimura, and K. Kuroda, Nature Photon. {\bf 6}, 662 (2012).
\bibitem{Hashimoto2017}  Y. Hashimoto, S. Daimon, R. Iguchi, Y. Oikawa, K. Shen, K. Sato, D. Bossini, Y. Tabuchi, T. Satoh,
B. Hillebrands, G. E. W. Bauer, T. H. Jo-hansen, A. Kirilyuk, T. Rasing, and E. Saitoh, Nature Commun. {\bf 8}, 15859 (2017).
\bibitem{Nemec2018}       P. \v Nemec, M. Fiebig, T. Kampfrath, and A. V. Kimel, Nature Phys. {\bf14}, 229 (2018).
%%%%%%%%%%%%%%% spin pumping in magnetic insulators   %%%%%%%%%%%%%%%Kajiwara2010,Heinrich2011,Ohnuma2014
\bibitem{Kajiwara2010} Y. Kajiwara, K. Harii, S. Takahashi, J. Ohe, K. Uchida, M. Mizuguchi, H. Umezawa, H. Kawai, 
K. Ando, K. Takanashi, S. Maekawa and E. Saitoh, Nature {\bf 464}, 262 (2010). 
\bibitem{Heinrich2011} B. Heinrich, C. Burrowes, E. Montoya, B. Kardasz, E. Girt, Young-Yeal Song, Yiyan Sun, and 
M. Wu, Phys. Rev. Lett. {\bf 107}, 066604 (2011).
\bibitem{Ohnuma2014} Y. Ohnuma, H. Adachi, E. Saitoh, and S. Maekawa, Phys. Rev. B {\bf 89}, 174417 (2014). 
%%%%%%%%%%%%%%%%%%%%%%%%%%%%%%%%%%%%%%%%%%%
\bibitem{Sturman1992}   B. I. Sturman and V. M. Fridkin, ``{\it The photovoltaic and photorefractive effects in noncentrosymmetric materials}'', (Gordon and Breach Science Publishers, 1992).
\bibitem{Tan2016} L. Z. Tan, F. Zheng, S. M. Young, F. Wang, S. Liu, and A. M. Rappe, NPJ Comput. Mater. {\bf 2}, 16026 (2016).
\bibitem{Tokura2018}    Y. Tokura and N. Nagaosa, Nature Commun. {\bf9}, 3740 (2018).
\bibitem{Belinicher1982}V. Belinicher, E. L. Ivcheriko, and B. Sturman, Zh. Eksp. Teor. Fiz. {\bf83}, 649 (1982).
\bibitem{Sipe2000}      J. E. Sipe and A. I. Shkrebtii, Phys. Rev. B {\bf61}, 5337 (2000)
\bibitem{Morimoto2016}  T. Morimoto and N. Nagaosa, Sci. Adv. {\bf 2}, e1501524 (2016).
\bibitem{Ishizuka2017c} H. Ishizuka and N. Nagaosa, New J. Phys. {\bf19}, 033015 (2017).
\bibitem{Nakamura2017}  M. Nakamura, S. Horiuchi, F. Kagawa, N. Ogawa, T. Kurumaji, Y. Tokura, and M. Kawasaki, Nature Commun. {\bf 8}, 281 (2017).
\bibitem{Ogawa2017}     N. Ogawa, M. Sotome, Y. Kaneko, M. Ogino, and Y. Tokura, Phys. Rev. B {\bf 96}, 241203 (2017).
%%%%%%%%%%%%%%  Staggered g factor   %%%%%%%%%%%%%%%%%%%
%Dender1997,Oshikawa1999,Feyerherm2000,Morisaki2007,Umegaki2009,Affleck1999
\bibitem{Dender1997} D. C. Dender, P. R. Hammar, D. H. Reich, C. Broholm, and G. Aeppli, Phys. Rev. Lett. {\bf 79}, 1750 (1997).
\bibitem{Oshikawa1999} M. Oshikawa, K. Ueda, H. Aoki, A. Ochiai, and M. Kohgi, J. Phys. Soc. Jpn. {\bf 68}, 3181 (1999).
\bibitem{Feyerherm2000} R. Feyerherm, S. Abens, D. G\"unther, T. Ishida, M. Meisner, M. Meschke, T. Nogami, and M. Steiner, 
J. Phys.: Condens. Matter {\bf 12}, 8495 (2000).
\bibitem{Morisaki2007} R. Morisaki, T. Ono, H. Tanaka, and H. Nojiri, J. Phys. Soc. Jpn. {\bf 76}, 063706 (2007). 
\bibitem{Umegaki2009} I. Umegaki, H. Tanaka, T. Ono, H. Uekusa, and H. Nojiri, Phys. Rev. B {\bf 79}, 184401 (2009). 
\bibitem{Affleck1999} I. Affleck and M. Oshikawa, Phys. Rev. B {\bf 60}, 1038 (1999); {\bf 62}, 9200 (2000).
%%%%%%%%%%%%%%%%%% Ising chains %%%%%%%%%%%%%%%%%%%%%%%%%
\bibitem{Shiba1980} N. Ishimura, and H. Shiba,  Prog. Theor. Phys. {\bf 63}, 743 (1980).
\bibitem{Coldea2010} R. Coldea, D. A. Tennant, E. M. Wheeler, E. Wawrzynska, D. Prabhakaran, M. Telling, K. Habicht, P. Smeibidl, K. Kiefer, Science {\bf 327} 177 (2010). 
\bibitem{Faure2018} Q. Faure, S. Takayoshi, S. Petit, V. Simonet, S. Raymond, L-P. Regnault, M. Boehm, 
J. S. White, M. M\^ansson, C. R\''uegg, P. Lejay, B. Canals, T. Lorenz, S. C. Furuya, T. Giamarchi and B. Grenier,  
Nature Phys. {\bf 14}, 716 (2018).
%%%%%%%%%%%%%%%%%% Chain MFT %%%%%%%%%%%%%%%%%%%%%%%%%Scalapino1975,Schulz1996,Sato2004,Okunishi2007
\bibitem{Scalapino1975} D. J. Scalapino, Y. Imry, and P. Pincus, Phys. Rev. B {\bf 11}, 2042 (1975).
\bibitem{Schulz1996}    H. J. Schulz, Phys. Rev. Lett. {\bf 77}, 2790 (1996).
\bibitem{Sato2004}      M. Sato and M. Oshikawa, Phys. Rev. B {\bf 69}, 054406 (2004).
\bibitem{Okunishi2007}  K. Okunishi and T. Suzuki, Phys. Rev. B {\bf 76}, 224411 (2007).
%%%%%%%%%%%%  JW transformation --1d many-body physics  %%%%%%%%%%%%Giamarchi2004,Gogolin2004,Tsvelik2007
\bibitem{Giamarchi2004} T. Giamarchi, {\it Quantum Physics in One Dimension}, (Oxford University Press, New York, 2003).
\bibitem{Gogolin2004}   A. O. Gogolin, A. A. Nersesian, and A. M.Tsvelik, {\it Bosonization and strongly correlated systems}, (Cambridge University Press, UK, 2004).
\bibitem{Tsvelik2007}   A. M. Tsvelik, {\it Quantum Field Theory in Condensed Matter Physics}, (Cambridge University Press, UK, 2007).

%\bibitem{Young2013}     S. M. Young, F. Zheng, and A. M. Rappe, Phys. Rev. Lett. {\bf 110}, 057201 (2013).
%\bibitem{Kim2017}       K.-W. Kim, T. Morimoto, and N. Nagaosa, Phys. Rev. B {\bf 95}, 035134 (2017).
%%%%%%%%%%%%%%%%%%%  multiferroics  iDM mechanism %%%%%%%%%%%%%%%%%%%%%%%%%%%
\bibitem{Katsura2005}     H. Katsura, N. Nagaosa, and A. V. Balatsky, Phys. Rev. Lett. {\bf 95}, 057205 (2005).
\bibitem{Takahashi2012}   Y. Takahashi, R. Shimano, Y. Kaneko, H. Murakawa, and Y. Tokura, Nature Phys. {\bf 8}, 121 (2012).
\bibitem{Huvonen2009}     D. H\''uvonen, U. Nagel, T. R\~o\~om, Y. J. Choi, C. L. Zhang, S. Park, and S.-W. Cheong, Phys. Rev. B {\bf 80}, 100402(R) (2009).
\bibitem{Furukawa2010}    S. Furukawa, M. Sato, and S. Onoda, Phys. Rev. Lett. {\bf 105}, 257205 (2010).
\bibitem{Tokura2014}      Y. Tokura, S. Seki, and N. Nagaosa, Rep. Prog. Phys. {\bf 77}, 076501 (2014).
\bibitem{Kraut1979}     W. Kraut and R. von Baltz, Phys. Rev. B {\bf 19}, 1548 (1979).
\bibitem{suppl}         Supplemental information.
%%%%%%%%%%%%%%%%%%%   multiferroics  magnetostriction  mechanism  %%%%%%%%%%%%%%%
%%%Pimenov2006,Miyahara2008,Mochizuki2010
\bibitem{Pimenov2006}     A. Pimenov, A. A. Mukhin, V. Yu. Ivanov, V. D. Travkin, A. M. Balbashov and A. Loidl, 
Nature Phys. {\bf 2}, 97 (2006).
\bibitem{Miyahara2008}    S. Miyahara and N. Furukawa, arXiv:0811.4082. 
\bibitem{Mochizuki2010-2} M. Mochizuki, N. Furukawa, and N. Nagaosa, Phys. Rev. Lett. {\bf 104}, 177206 (2010). 
\bibitem{Katsura2009}     H. Katsura, M. Sato, T. Furuta, and N. Nagaosa, Phys. Rev. Lett. {\bf 103}, 177402 (2009).
%%%%%%%%%%%%%%%%%%
\bibitem{Hirobe2017}    D. Hirobe, M. Sato, T. Kawamata, Y. Shiomi, K. Uchida, R. Iguchi, Y. Koike, S. Maekawa, and E. Saitoh, Nature Phys. {\bf13}, 30 (2017).
\bibitem{Jia2006}       C. Jia, S. Onoda, N. Nagaosa, and J. H. Han, Phys. Rev. B {\bf 74}, 224444 (2006).
\bibitem{Ishizuka2018}  H. Ishizuka and N. Nagaosa, preprint (arXiv:1806.06833) (2018).
\end{thebibliography}
\end{document}

\preprint{APS/123-QED}

\title{
  Supplemental Information for \\
``{\it Nonlinear Optical Rectification of Spin Current: Magnetic Shift Current}''
}

\author{Hiroaki Ishizuka}
\affiliation{
Department of Applied Physics, The University of Tokyo, Bunkyo, Tokyo, 113-8656, JAPAN 
}

\author{Masahiro Sato}
\affiliation{
Department of Physics, Ibaraki University,  Mito, Ibaraki, 310-8512, JAPAN
}

\date{\today}

% \begin{abstract}
% \end{abstract}

\pacs{
}% PACS, the Physics and Astronomy
% Classification Scheme.

\maketitle
\onecolumngrid
%%%%%%%%%%%%%%%%%%%%%%%%%%%%%%%%%%%%%%%%%%%%%
%%%%%%%%%%%%%%%%%%%%%%%%%%%%%%%%%%%%%%%%%%%%%
\section{Symmetry property of the nonlinear spin conductivity} 

In this section, we discuss the general form of the quadratic spin conductivity $\sigma^{(2)}(\omega)$ based on symmetry. 
The symmetry argument enables us to restrict the possible from of $\sigma^{(2)}(\omega)$.  
For clarity, the parameter dependence of $\sigma^{(2)}$ is explicitly shown in this section, i.e., 
$\sigma^\text{(2)}(\omega)=\sigma^\text{(2)}(\omega;\delta,h_s,\cdots)$, where $\cdots$ are the coupling constants 
of the spin-electromagnetic-field couplings. 
By definition, the relation between the spin current and an external electromagnetic field is given by 
\begin{align}
\langle \hat J_{sc}(\delta)\rangle_{\delta,h_s,\cdots}\approx\sigma^{(2)}(\delta,h_s,\cdots)C^2.
\label{eq:suppl:defsigma}
\end{align}
Here, $\hat J_{sc}(\delta)$ is the spin current operator, $\langle \hat A\rangle_{\delta,h_s,\cdots}$ 
is the thermal average of an observable $\hat A$, and $C$ is the magnitude of the external oscillating electromagnetic field; 
in case of Zeeman coupling, $C$ is the magnitude of the magnetic field $B(t)\equiv C \cos(\omega t)$, 
while it is the magnitude of the electric field $E(t)\equiv C \cos(\omega t)$ in the system with inverse DM or magnetostriction coupling. 
Equation~\eqref{eq:suppl:defsigma} will be used to study the symmetry property 
of the conductivity $\sigma^{(2)}(\omega)$ below.

We first discuss the symmetries of the spin-chain system 
with a Zeeman coupling $H_\text{Z}(t)=-B(t)\sum_i(\eta-(-1)^i\eta_s)S_i^z$. 
Let us first consider the site-centered inversion operation, $\bm S_i\to\bm S_{-i}$. Acting this transformation to the system, 
we find $\hat H(\delta,h_s)\to\hat H(-\delta,h_s)$, $\hat J_{sc}(\delta)\to-\hat J_{sc}(-\delta)$, $\hat H_\text{Z}(\eta_s)\to-\hat H_\text{Z}(\eta_s)$, and $C\to C$. Therefore, the thermal average of spin current satisfies the following equality:
\begin{align}
\langle \hat J_{sc}(\delta)\rangle_{\delta,h_s,\eta_s}
=
-\langle \hat J_{sc}(-\delta)\rangle_{-\delta,h_s,\eta_s}.\label{suppl:eq:defsigma1}
\end{align}
On the other hand, the definition of the conductivity in Eq.~\eqref{eq:suppl:defsigma} gives 
 \begin{align}
\langle \hat J_{sc}(\delta)\rangle_{\delta,h_s,\eta_s}&\equiv\sigma^{(2)}(\delta,h_s,\eta_s)C^2,\nonumber\\
\langle \hat J_{sc}(-\delta)\rangle_{-\delta,h_s,\eta_s}&\equiv\sigma^{(2)}(-\delta,h_s,\eta_s)C^2.\label{suppl:eq:defsigma2}
\end{align} 
From Eqs.~\ref{suppl:eq:defsigma1} and \ref{suppl:eq:defsigma2}, we find
\begin{align}
\sigma^{(2)}(\delta,h_s,\eta_s)=-\sigma^{(2)}(-\delta,h_s,\eta_s).\label{eq:suppl:Zeeman_site}
\end{align}
Next, we consider the inversion operation at a bond center, $\bm S_i\to\bm S_{1-i}$. Similar to the site-center-inversion operation, we find $\hat H(\delta,h_s)\to\hat H(\delta,-h_s)$, $\hat J(\delta)\to-\hat J(\delta)$, $\hat H_\text{Z}(\eta_s)\to\hat H_\text{Z}(-\eta_s)$, and $C\to C$. From these symmetry relations, we obtain
\begin{align}
\sigma^{(2)}(\delta,h_s,\eta_s)C^2\equiv\langle \hat J_{sc}(\delta)\rangle_{\delta,h_s,p'}
=-\langle \hat J_{sc}(\delta)\rangle_{\delta,-h_s,-\eta_s}\equiv-\sigma^{(2)}(\delta,-h_s,-\eta_s)C^2,
\end{align}
and
\begin{align}
\sigma^{(2)}(\delta,h_s,\eta_s)=-\sigma^{(2)}(\delta,-h_s,-\eta_s).
\label{eq:suppl:Zeeman_bond}
\end{align}
One of the simplest form that satisfies Eqs.~\eqref{eq:suppl:Zeeman_site} and \eqref{eq:suppl:Zeeman_bond} is $\sigma^{(2)}(\delta,h_s,\eta_s)\propto \delta\,h_s \eta_s^2$. This is the result in Eq.~\eqref{eq:sigma_Zeeman} of the main text. 

In case of the inverse DM coupling $H_\text{iDM}$, we find
\begin{align}
\sigma^{(2)}(\delta,h_s,p,p_s)=-\sigma^{(2)}(-\delta,h_s,p,-p_s),\label{eq:suppl:iDM_site}
\end{align}
for the site-centered inversion, and
\begin{align}
\sigma^{(2)}(\delta,h_s,p,p_s)=-\sigma^{(2)}(\delta,-h_s,p,p_s),\label{eq:suppl:iDM_bond}
\end{align}
for the bond-centered inversion. Therefore, the simplest leading terms of $\sigma^{(2)}$ are given 
by $\sigma^{(2)}\propto h_s p_s$ or $h_s\delta$. 
This form agrees with that found in Eq.~\eqref{eq:sigmaAS}. 

In the system with magnetostriction coupling $H_\text{ms}$, the same symmetry argument leads to the relation 
\begin{align}
\sigma^{(2)}(\delta,h_s,A,A_s)=-\sigma^{(2)}(-\delta,h_s,A,-A_s),\label{eq:suppl:MS_site}
\end{align}
for the site-centered inversion, and
\begin{align}
\sigma^{(2)}(\delta,h_s,A,A_s)=-\sigma^{(2)}(\delta,-h_s,A,A_s),\label{eq:suppl:MS_bond}
\end{align}
for the bond-centered inversion. The result in Eq.~\eqref{eq:sigma_ms} in the main text 
is consistent with these symmetry-related properties.

%%%%%%%%%%%%%%%%%%%%%%%%%%%%%%%%%%%%%%%%%%%%%
%%%%%%%%%%%%%%%%%%%%%%%%%%%%%%%%%%%%%%%%%%%%%
%%%%%%%%%%%%%%%%%%%%%%%%%%%%%%%%%%%%%%%%%%%%%
%%%%%%%%%%%%%%%%%%%%%%%%%%%%%%%%%%%%%%%%%%%%%
\section{Estimate of the required electromagnetic field}

In this section, 
we elaborate on the order estimate of the required magnitude of the external oscillating electromagnetic field. 
We use the experimental results of the spin Seebeck effect as the standard of spin current magnitude.

%%%%%%%%%%%%%%%%%%%%%%%%%%%%%%%%%%%%%%%%%%%%%
%%%%%%%%%%%%%%%%%%%%%%%%%%%%%%%%%%%%%%%%%%%%%
\subsection{Criterion for the experimentally observable spin current}

In this subsection, we define the criteria for the observable spin current by using established results of 
spin Seebeck effect, i.e., the spin current induced by thermal gradient~\cite{Uchida2008,Hirobe2017}. 
As a simple system, we focus on the spin Seebeck effect in ferromagnetic insulators~\cite{Uchida2008}. 
The typical magnitude of the spin current will be estimated using semiclassical Boltzmann theory below.

A three-dimensional (3D) spin-$S$ Heisenberg ferromagnet with a uniaxial spin anisotropy is considered. 
The Hamiltonian reads:
\begin{align}
H_\text{Heis}=-J_H\sum_{\langle i,j\rangle}\bm S_i\cdot\bm S_j-D\sum_i (S_i^z)^2-h\sum_i S_i^z.
\end{align} 
Here, $J_H>0$ is the ferromagnetic exchange interaction between the neighboring spins, $D$ is the uniaxial anisotropy, and $h$ is the external magnetic field. The magnon dispersion of this model reads
\begin{align}
\varepsilon(\bm k)=2J_HS\left(3-\sum_{a=x,y,z}\cos(k_a a)\right)+2DS+h\approx J_HSk^2+2DS+h,
\end{align}
and the magnon group velocity along the $a$ axis is
\begin{align}
v_a(\bm k)=\frac{2J_HS a}\hbar\sin(k_a a)\approx \frac{2J_HS a^2}\hbar k_a.
\end{align}
This free boson model well describes the low-energy physics of the ferromagnet below the critical temperature.

A semiclassical Boltzmann theory is used to calculate the magnon spin current. The spin current along the $z$ axis reads
\begin{align}
J_{sc}(\bm r)=\hbar\int \frac{d\bm k}{(2\pi)^3} v_z f_{\bm k}(\bm r). 
\end{align}
Here, $\hbar$ is the Planck constant and $f_{\bm k}(\bm r)$ is the magnon density 
with momentum $\bm k$ at position $\bm r$. When a temperature gradient is applied to the ferromagnet, 
the magnon density in the non-equilibrium steady state is determined from the Boltzmann transport equation:
\begin{align}
\bm v_{\bm k}\cdot\bm \nabla_r f_{\bm k}(\bm r)=-\frac{f_{\bm k}(\bm r)-f^{(0)}_{\bm k}(\bm r)}{\tau_{\bm k}}.
\end{align}
Here, $\tau_{\bm k}=\hbar/[\alpha \varepsilon(\bm k)]$ is the magnon relaxation time, $f_{\bm k}(\bm r)$ is the magnon distribution in the nonequilibrium steady state, and 
$f^{(0)}_{\bm k}(\bm r)\equiv 1/(\exp[\beta(\bm r)\varepsilon(\bm k)]-1)$ is the equilibrium Bose distribution function; $\beta(\rm r)=1/k_BT(\bm r)$ is the inverse temperature where $T(\bm r)$ is the temperature at $\bm r$ and $k_B$ is Boltzmann constant.
The parameter $\alpha$ in $\tau_{\bm k}$ is the Gilbert damping constant, which is a dimensionless material-dependent parameter; typically, it is $\alpha=0.1-0.001$~\cite{Makino2012,Kehlberger2015,Tang2018}. 
%We assumed $f_{\bm k}(\bm r)$ does not change over time, because we focus on the steady state. 
Assuming $f_{\bm k}(\bm r)$ is sufficiently close to $f^{(0)}_{\bm k}(\bm r)$, we may rewrite $f_{\bm k}(\bm r)$ as 
\begin{align}
f_{\bm k}(\bm r)=f^{(0)}_{\bm k}(\bm r)+g_{\bm k}(\bm r).
\end{align}
To the leading order in $g_{\bm k}(\bm r)$, the Boltzmann equation becomes
\begin{align}
\bm v_{\bm k}\cdot\bm \nabla_r f^{(0)}_{\bm k}(\bm r)=-\frac{g_{\bm k}(\bm r)}{\tau_{\bm k}}.
\end{align}
Therefore, $g_{\bm k}(\bm r)$ is given by 
\begin{align}
g_{\bm k}(\bm r)=&
\frac{\hbar\beta(\bm r) e^{\beta(\bm r)\varepsilon_{\bf k}}}{\alpha(e^{\beta(\bm r)\varepsilon_{\bf k}}-1)^2} 
\bm v(\bm k)\cdot\frac{\bm \nabla_r T(\bm r)}{T(\bm r)}.
\end{align}
Assuming a simple, linear temperature gradient along the $z$ axis, $T(\bm r)=T_0+\Delta T z$, 
we have 
\begin{align}
g_{\bm k}(\bm r)=
\frac{\hbar\beta(\bm r) e^{\beta(\bm r)\varepsilon_{\bf k}}}{\alpha(e^{\beta(\bm r)\varepsilon_{\bf k}}-1)^2} v_z(\bm k)\frac{\Delta T}{T(\bm r)}.
\end{align}
The spin current density hence reads
\begin{align}
J_{sc}(\bm r)=\frac{\hbar^2}{\alpha k_B T(\bm r)} \int_{BZ} \frac{d\bm k}{(2\pi)^3} v_z^2(\bm k)\frac{ e^{\beta(\bm r)\varepsilon_{\bf k}}}{(e^{\beta(\bm r)\varepsilon_{\bf k}}-1)^2}\frac{\Delta T}{T(\bm r)}. \label{eq:J3dFM}
\end{align}
Here, the integral is over the first Brillouin zone.

The integral in Eq.\eqref{eq:J3dFM} is estimated using approximations $v_z(\bm k)\approx (2J_Ha^2/\hbar)k_z$ and 
$\varepsilon_{\bf k}\approx J_Ha^2k^2+2D+h$. 
Approximating the Brillouin zone by a globe of the same volume, we find
\begin{align}
J_{sc}(\bm r)\approx&\frac{3(6\pi^2)^{\frac23}J_H^2S^2}{2\alpha k_B a T(\bm r)}
\frac{\Delta T}{T(\bm r)}
F\left(\frac{J_HSa^2\Lambda^2}{2k_B T(\bm r)},\frac{2DS+h}{2k_B T(\bm r)}\right).
\end{align}
Here, $\Lambda=(6\pi^2)^{1/3}/a$ and
\begin{align}
F(a,b)= \int_0^1x^4{\rm csch}^2(ax^2+b)dx.
\end{align}
Using a set of typical parameters $S=1$, $J_H=100k_B$ J, $D=0$ J, $h=\mu_B$ J, $a=4\times10^{-10}$ m, $\alpha=10^{-2}$, $T=100$ K, $\Delta T=3\times10^4$ K/m, the spin current density is $J_{sc}^\text{(FM)}\sim 10^{-12}$ J/cm$^2$.

It is generally difficult to directly observe spin current since the electron spin is a chargeless quantity. 
Instead of the direct measurement, the usual experimental set up of spin Seebeck effect indirectly measures 
the thermally generated spin current by changing it into electric current/voltage 
through the inverse spin Hall effect~\cite{Uchida2008}.    
The typical spin-Seebeck voltage for the ferromagnet is about $\sim10$ $\mu$V. 
Therefore, it is natural to assume that the estimated typical value of spin current 
$J_{sc}^\text{(FM)}\sim 10^{-12}$ J/cm$^2$ corresponds to the typical value of the voltage $\sim10$ $\mu$V. On the other hand, the spin-Seebeck voltage with the order of $\sim 1$ nV is observed in a recent experiment of 
the spinon spin Seebeck effect in a quantum spin chain magnet $\rm Sr_2RuO_3$~\cite{Hirobe2017}. 
Namely, the magnon spin current is about $10^4$ times as large as the spinon spin current. 
This is also supported by the recent theoretical calculation~\cite{Hirobe2017}. 
Therefore, in order to determine the required electromagnetic field in our spin chain model, 
it would be reasonable to use the value of spin current $J_{sc}^{(0)}=J_{sc}^\text{(FM)}\times 10^{-4}\sim 10^{-16}$ J/cm$^2$ as the standard. In fact, the value of 1 nV observed in Ref.~\onlinecite{Hirobe2017} is quite close to the lowest limit of experimentally resolved voltage. 

%%%%%%%%%%%%%%%%%%%%%%%%%%%%%%%%%%%%%%%%%%%%%%
\subsection{Required electromagnetic field for an observable magnetic shift current in the spin chain model}

We consider a quantum spin chain model with $J\sim10^2k_B$ J, $\delta\sim0.1$, $h_s\sim 0.1\times\mu_B$ J, 
and lattice parameter $a_0\sim4$ \AA  to estimate the magnitude of photo-induced spin current. 
Here, $k_B$ is Boltzmann constant and $\mu_B$ is Bohr magneton. 
The spin chain has a spinon band gap of $\Delta_g\sim 10^{-22}$ J. 
Hence the frequency of the external electromagnetic field is set to be $\omega\sim 10^{-22}$ J, i.e., the frequency range from gigahertz to terahertz. These values are typical for transition metal magnets. In this section, we focus on the zero temperature limit.

For the case of Zeeman coupling $H_\text{Z}$, we assume $\eta_s \sim 0.1\times \mu_B = 9.274\times10^{-25}$ J/T. 
By substituting these values into Eq.~\eqref{eq:sigma_Zeeman}, we find $\sigma^{(2)}\sim 10^{-28}$ J/T$^2$ 
per a chain.
Considering a bundle of the spin chains with $a_0$ distance, we find the conductivity to be
\begin{align}
\sigma^{(2)}_{3D}\equiv \sigma^{(2)}/a_0^2 \sim 10^{-9}\,\text{J/T$^2$m$^2$}.
\end{align}
By definition, the spin current reads $J_{sc}=\sigma^{(2)}B^2$. 
The required magnetic flux density $B$ for a given $J_{sc}$ thereby reads
\begin{align}
B=\sqrt{J_{sc}/\sigma^{(2)}_{3D}}.
\end{align}
Substituting $J_{sc}=J_{sc}^{(0)}=10^{-12}$ A/m$^2$ and the $\sigma^{(2)}_{3D}$ above, 
we find $B\sim 10^{-2}$ T. 
The corresponding electric field strength is $E=c B\sim 10^4$ V/cm. 
Here, $c\sim10^8$ m/s is the speed of light and we assume the relative electric permittivity $\epsilon_r=10$. 
The electric field $10^3$ V/cm is experimentally accessible, and 
hence the spin current induced by the Zeeman coupling  would be observable in real experiments. 

We next consider the contribution from the inverse DM effect. 
The magnitude of coupling constants for the inverse DM effect $p$ and $p_s$ are estimated 
in a recent study on oxides~\cite{Jia2006}. The estimated polarization density reads $P\sim10^2$ nC/cm$^2$; 
the result gives $P a_0^3\sim 10^{-31}$ Cm per a bond~\cite{Ishizuka2018}. We use $p\sim 10p_s\sim 10^{-32}$ Cm 
for the estimate. Substituting these values, we find $\sigma^{(2)}\sim 10^{-44}$ A$^2$s$^4$/kg, and 
$\sigma^{(2)}_{3D}\equiv \sigma^{(2)}/a_0^2 \sim 10^{-25}$ A$^2$s$^4$/m$^2$kg. 
Therefore, $E\sim 10^5$ V/cm is required for $J_{sc}^{(0)}=10^{-12}$ A/m$^2$.

In the last, we consider the contribution from the magnetostriction effect. 
In the insulating magnets, the magnetostriction effect comes from the lattice distortion (i.e., phonon) induced 
by an external electric field~\cite{Tokura2014}. We simply focus on the $q=\pi$ mode of lattice distortion 
in a 1D phonon system. Within the Landau-theory framework, the free energy of a phonon system coupled to 
an external electric field $E$ is given by
\begin{align}
F=\varepsilon_{ph}(q)u_q^2-q_eEa_0u_q.
\end{align}
Here, $\varepsilon_{ph}(q)$ is the phonon energy of the wave number $q$, $q_e$ is the charge of magnetic ions, 
and $u_q$ is the distortion with wave number $q$. By minimizing the free energy, $u_q$ becomes
\begin{align}
u_q(E)=\frac{q_eEa_0}{2\varepsilon_{ph}(q)}.
\end{align}
The effect of this modulation on the exchange interaction is estimated from an assumption
\begin{align}
J_{i,i+1}(u_{i+1}-u_i)=J_{i,i+1}(0)e^{-\zeta (u_{i+1}-u_i)},
\end{align}
where $\zeta$ is a dimensionless constant of ${\cal O}(1)$. 
This corresponds to the tight-binding approximation, i.e., the wave function is bounded to the ions 
with exponentially decaying tails. 
With these assumptions, the modulation of the exchange coupling reads:
\begin{align}
J_{i,i+1}(u_{i+1}-u_i)\sim J_{i,i+1}(0)(1-\zeta (u_{i+1}-u_i)).
\end{align}
We hence obtain the coupling constant of magnetostriction term as 
\begin{align}
A_s\sim \frac{Jq_ea_0}{2\varepsilon_{ph}(\pi)}.
\end{align}
Using $J\sim 10^2k_B \text{J}$, $q=1.6\times10^{-19} \text{C}$, $a_0\sim 4\times10^{-10}\text{m}$, 
and $\varepsilon_{ph}(\pi)\sim k_B\Theta_D\sim 10^3k_B$, we find $A_s\sim10^{-28}$ Jm/V. 
Using this value for the magnetostriction coupling, we find $\sigma^{(2)}\sim 10^{-38}$ A$^2$s$^4$/kg and 
$\sigma^{(2)}_{3D}\equiv \sigma^{(2)}/a_0^2 \sim 10^{-19}$ A$^2$s$^4$/m$^2$kg. 
Therefore, $E=\epsilon_r[J_{sc}^{(0)}/\sigma^{(2)}_{3D}]^{1/2}\sim 10^2$ V/cm is required 
for the standard value of the spin current $J_{sc}^{(0)}=10^{-12}$ A/m$^2$. 
Here, $E$ is renormalized by $\epsilon_r$ as the electric field is partially screened in materials. 
As $J_{sc}\propto E^2$, the magnitude of $J_{sc}$ induced by magnetostriction effect 
is about two orders of magnitude larger than that by the Zeeman coupling.

%%%%%%%%%%%%%%%%%%%%%%%%%%%%%%%%%%%%%%%%%%%%%%